\documentclass[openacc]{rstransa}

\begin{document}

\title{Mass loss and stellar superwinds}

\author{Jorick S. Vink$^{1}$}

\address{$^{1}$Armagh Observatory, College Hill, Armagh BT61 9DG, Northern Ireland}



\corres{Jorick S. Vink\\
\email{jsv@arm.ac.uk}}

\begin{abstract}
Mass loss bridges the gap between massive stars and supernovae (SNe) in 
two major ways: (i) theoretically it is the amount of mass lost that determines 
the mass of the star prior to explosion, and (ii) observations of the 
circumstellar material around SNe may teach us the type of progenitor
that made the SN. Here, I present the latest models and observations 
of mass loss from massive stars, both for canonical massive O stars, as well
as very massive stars (VMS) that show Wolf-Rayet type features.  
\end{abstract}


\begin{fmtext}

\section{Introduction: massive stars with superwinds before explosion}

The progenitor stars of Type IIP(lateau) Supernovae (SNe) have been established to 
be red supergiants (RSGs) (Van Dyk et al. 2002; Smartt et al. 2009), but the 
more massive progenitors of Type IIb, IIn, Ibn, Ibc have yet to be identified.
Traditionally, SN progenitors have been pinpointed using photometry, but since 
recently a new method of rapid ``flash'' spectroscopy has become available:
Gal-Yam et al. (2014) discovered a stellar wind in the SN spectrum
of the Type IIb supernova 2013cu, which they attributed to be a Wolf-Rayet (WR)
star due to the strong emission lines in its spectrum (see Fig.\,1).

However, Groh (2014) and Gr\"afener \& Vink (2016) modelled the spectrum, 
arguing the narrow lines from 2013cu are
representative of a luminous blue variable (LBV) or 
other post-RSG object.  Gr\"afener \& Vink (2016) derived a huge mass loss prior to explosion, 
arguing the progenitor had been subjected to a {\it superwind}, due to 
its increasing radiation pressure and associated 
Eddington factor $\Gamma_{\rm e} = g_{\rm rad}/g_{\rm grav}=~\sigma_{\rm e} L/(4 \pi c G M)$ 
prior to explosion (see Fig.\,2) 

It is clear that we need to understand 
the mass-loss properties of massive stars at high $\Gamma_{\rm e}$, but before addressing 
this aspect we first review the mass-loss rates of normal O-stars (at moderate 
$\Gamma_{\rm e}$ parameters). I assume basic knowledge of radiation-driven 
wind theory, but see Owocki (2015) and Vink (2015) for more information.

\end{fmtext}

\maketitle

\begin{figure*}
\begin{center}
\includegraphics[width=\textwidth]{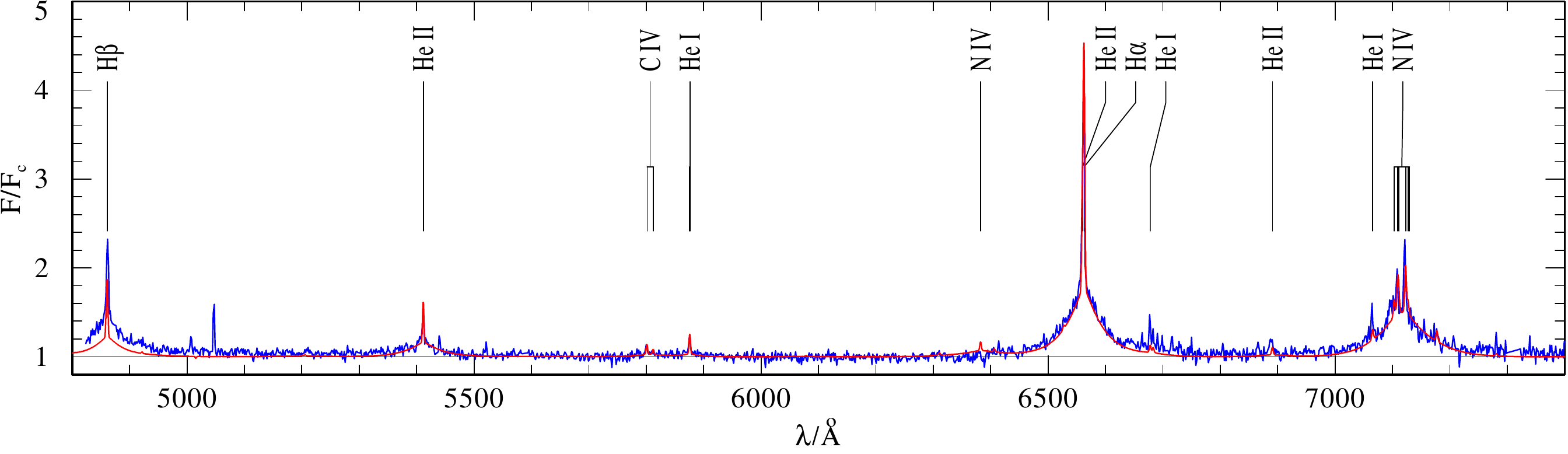}
\caption{Model fit to the observed spectrum of the IIb supernova SN 2013cu, 15.5 hours after 
explosion (Gal-Yam et al. 2014). See Gr\"{a}fener \& Vink (2016) for more details on the model fit.}
\label{fig_gal}
\end{center}
\end{figure*}

\begin{figure*}
\begin{center}
\includegraphics[width=\textwidth]{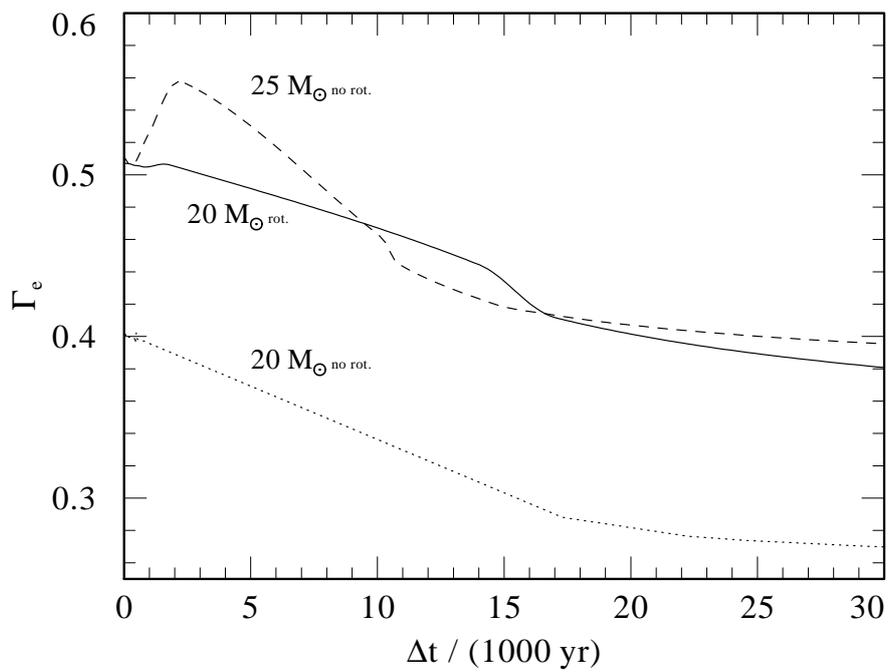}
\caption{Eddington factor in the final thousands of years before expiration. 
See Gr\"afener \& Vink (2016) with evolutionary model data from Ekstr\"om et al. (2012).}
\label{fig_ekstrom}
\end{center}
\end{figure*}

\vspace*{-5pt}

\section{Normal O star wind models}

The radiation-driven wind theory was developed in the 1970s by
Lucy \& Solomon (1970) and Castor, Abbott \& Klein (1975; hereafter CAK).
For many years it was thought that the strong, optically thick, 
lines of carbon, oxygen and nitrogen (CNO) were the most important wind drivers -- due 
to the relatively high abundance of these
elements -- causing their presence in the ultraviolet (UV) spectra of 
early-type stars. 

However, we now know that despite its relatively minor cosmic abundance, it is
the millions of weak (and optically thin) iron (Fe) lines below the critical (sonic) point 
that set the mass-loss rates in massive O-star winds (Vink et al. 1999, 2000, 2001).
The mass-loss recipes from these authors are now widely used in the 
evolution of massive stars, but questions 
have been raised regarding the absolute strengths of the normal\footnote{Furthermore, at low 
luminosities, with $\log(L/\L_{\odot}) < 5.2$ there is a {\it weak wind} problem (Martins et al. 2005, Puls et al. 2008). 
This corresponds to spectral types later than O6.5V (Muijres et al. 2012) 
and may be due to a lack of subsonic radiative driving due to 
a recombination of Fe {\sc v} to {\sc iv} at that $T_{\rm eff}$. If true, this would 
imply the physics of an ``inverse'' bi-stability jump} O-star winds, due 
to the widespread presence of wind inhomogeneities, or clumping (e.g. 
Puls et al. 2008; Cantiello et al. 2009).

\section{Normal O star wind observations}

Fullerton et al. (2006) highlighted the problem of wind clumping in the 
form of the so-called Phosphorus {\sc v} problem: if the winds of massive 
stars are inhomogeneous and the clumps can be treated in the micro-clumping 
approximation (see also Bouret et al. 2003), then the most likely solution for the P {\sc v} problem in the UV
part of the spectrum would be {\it far} lower mass-loss rates (see Fig.\,3), perhaps by 
an order-of-magnitude.

\begin{figure*}
\begin{center}
\includegraphics
  [width=\textwidth]{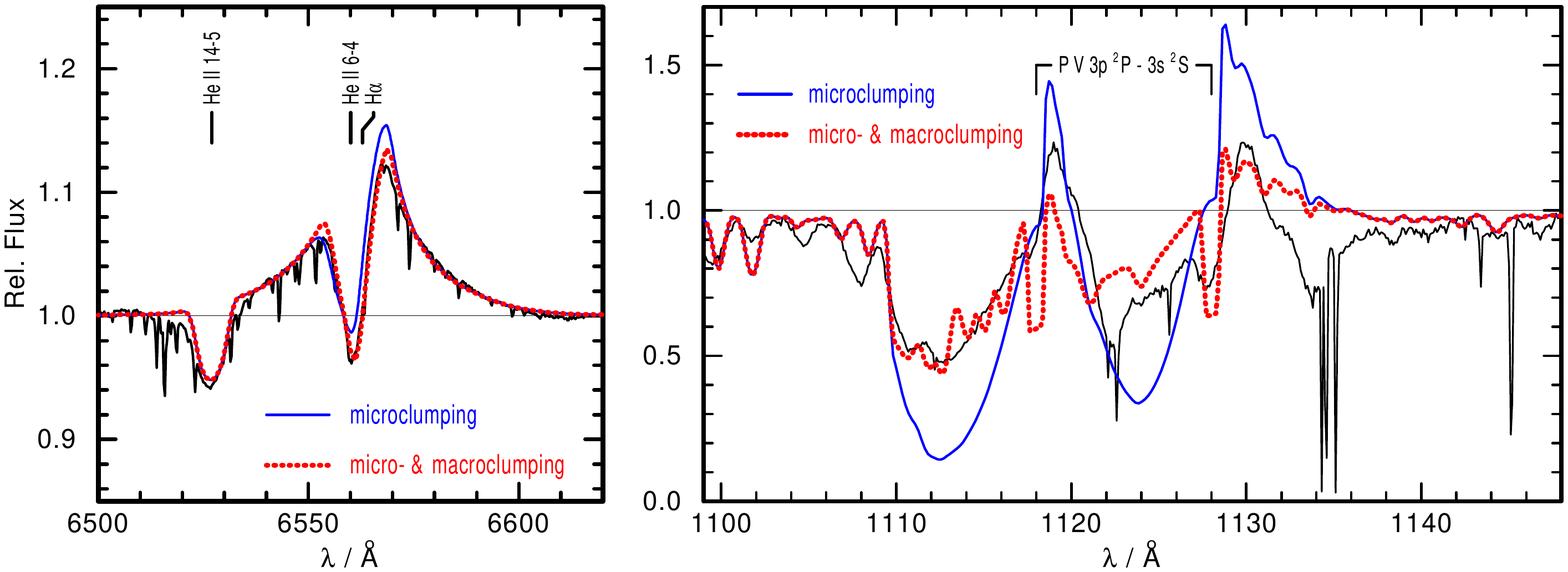}
\caption{The phosphorus {\sc v} problem, as solved by Oskinova et al. (2007) using 
optically thick ``macro'' clumps. 
The figure on the right is the ultraviolet P {\sc v} spectrum of the 
O supergiant $\zeta$ Pup, whilst the figure on the left concerns the H$\alpha$ profile.}
\end{center}
\end{figure*}

The P {\sc v} problem was solved when Oskinova et al. (2007) (see also Sundqvist et al. 2010; Surlan et al. 2013) 
introduced the concept of porosity or ``macro-clumping'', previously constructed for X-ray observations, to 
the UV part of the spectrum (see Fig.\,3). The P{\sc v} line can now be fit, without
needing to reduce the empirical mass-loss rate excessively. The unclumped empirical mass-loss rates
are still overestimated by a factor of the square-root of the clumping factor, but as long as clumping is moderate, with clumping
factors 6-8, the reduced mass-loss rates (by factor 2-3) agree with the theoretical rates of Vink et al. 
(see the discussions in Repolust et al. 2004; Mokiem et al. 2007; Ramirez-Agudelo et al. 2016).

\section{The first mass-loss calibration}

For normal O-type stars, say in the luminosity range $\log(L/\L_{\odot}) = 5.2 - 5.8$, there is no guarantee that 
the clumping factor is a factor 6-8, as may seemingly have been suggested in the previous paragraph. 
If the {\it true} clumping factor is larger than 6-8, the Vink et al. rates would be too large compared to empirical values.
The interesting question is now if we can establish what the correct clumping factor is in O-star winds?

For the more extreme winds of Wolf-Rayet stars (including both classical WR stars, as well as very massive stars (VMS)
of the WNh sequence) the spectral lines are in emission, and owing to the presence of electron scattering wings, which depend
linearly on density (versus the quadratic density behaviour of the central recombination part of the emission line), 
one may constrain the clumping factor (Hillier et all. 1991). On the basis of this methodology the Potsdam group
(e.g. Hamann \& Koesterke 1998) usually quote clumping factors in the range 4-16, with mass-loss rate reductions 
for WR stars (in comparison to unclumped empirical rates) of $\sim2-4$, 
with a reasonable average clumping factor of $\sim$10, i.e. a mass-loss rate reduction of $\sim$3 (Hamann et al. 2008).

For decades it was known that Wolf-Rayet stars are special in comparison to O stars in that the
wind efficiency number $\eta = \dot{M}v_{\infty}/(L/c)$ exceeds unity due to highly efficient multiple scattering 
(e.g. Gayley et al. 1995). 
However, it was not until it became clear that VMS with masses exceeding $150M_{\odot}$ exist (Crowther et al. 2010, Bestenlehner et al. 2011, 
Vink et al. 2015) that we were able to construct a mass (and luminosity) sequence of O and VMS-WNh stars, sampling the 
entire upper initial mass function (IMF), including 
the entire O-Of-Of/WN-WNh spectral sequence -- and to utilize this to calibrate 
stellar wind mass-loss rates using the $\eta = \tau = 1$ relation (Vink \& Gr\"afener 2012).

Vink \& Gr\"afener (2012) showed that the transition luminosity between optically thin Of stars (with $\tau < 1$) and 
optically thick (with $\tau > 1$) WNh stars in the Arches cluster
(based on CMFGEN modelling by Martins et al. 2008) could be utilised to derive 
the transition mass-loss rate. As this model-independent 
transition mass-loss rate agrees well with both the theoretical mass-loss rates of 
Vink et al., as well as the empirical mass-loss rates of Martins et al. (for an assumed clumping factor of 10) this 
provides a strong consistency argument -- resulting in the first mass-loss calibration in astronomy.

Moreover, this provides evidence for strong mass-loss for VMS, ensuring ``evaporation'' of Galactic stars due to mass loss. 
This would make it highly unlikely that exotic phenomena involving 
gamma-ray bursts (Vink \& de Koter 2005), pair-instability SNe (Langer et al. 2007), or other super-luminous SNe 
(Quimby et al. 2011; Neill et al. 2011; Stoll et al. 2011; Lunnan et al. 201; Leloudas et al. 2015; Vreeswijk et al. 2015; 
Chen et al. 2016)
would take place at Galactic ``high'' metal content, but that low metallicity may be a prerequisite. 

Moreover, the very existence 
of relatively massive stellar black holes, with masses $> 30M_{\odot}$ 
as recently discovered through gravitational waves (Abbott et al. 2016) would 
call for low-metallicity environments according to our wind models (Belczynksi et al. 2010) -- 
providing a relationship between the maximum mass of stellar black holes and metallicity.

\section{Very Massive Stars with high Eddington factor}

\begin{figure*}
\begin{center}
\includegraphics
  [width=\textwidth]{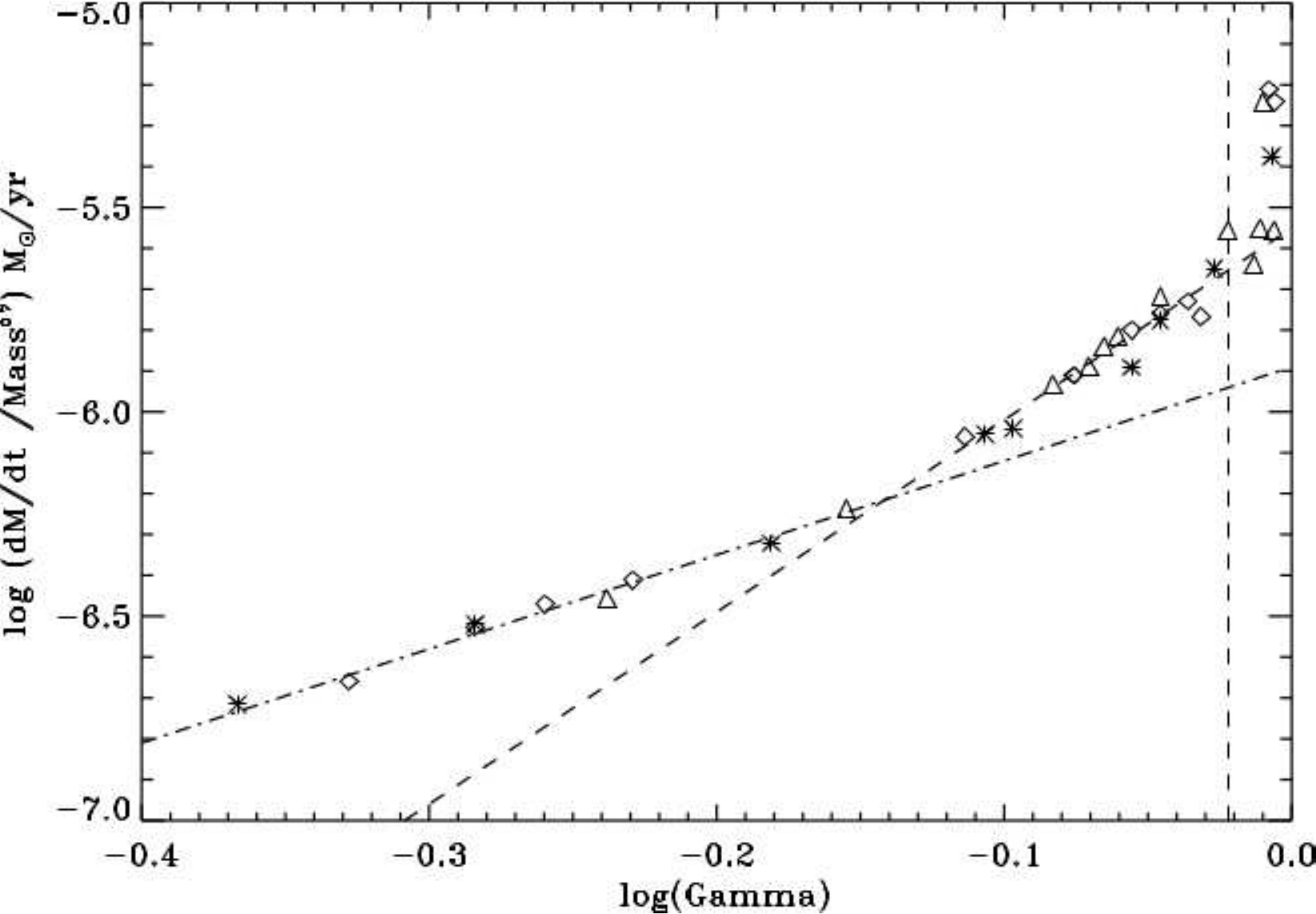}
\caption{VMS mass-loss rates versus the Eddington factor, as predicted by Monte Carlo
models of Vink et al. (2011) using a new parametrisation of the line acceleration by 
M\"uller \& Vink (2008).}
\end{center}
\end{figure*}

The previous section concluded that we know the mass-loss rates of massive stars at the 
transition luminosity very accurately. As long as the clumping factor for lower luminosity
``normal'' O-stars in the luminosity range $\log(L/\L_{\odot}) = 5.2 - 5.8$ is not much larger
than $\sim$10, we can be confident that the Vink et al. (2000) relation provides accurate mass-loss 
rates, as these rates have now been calibrated at the high mass-luminosity end, against a model
independent anchor (which is clumping and porosity independent). 

However, for VMS with luminosities {\it above} the transition luminosity, the Vink et al. relationship
would {\it under}estimate the true mass-loss rates, as the Vink et al. (2000) relation
was derived for optically thin winds, whilst VMS have winds that are optically thick (see also 
Gr\"afener \& Hamann 2008; Gr\"afener et al. 2011). 

In order to study the transition from optically thin to optically thick winds, Vink et al. (2011) 
computed Monte Carlo radiative acceleration  models (based on an original concept by Abbott \& Lucy 1985) into
the regime of VMS up to 300$M_{\odot}$. The results are shown in Fig.\,4. The plot shows a relatively 
shallow slope for optically thin O-star winds with $\dot{M} \propto \Gamma^2$ (in reasonable agreement 
with CAK theory) at relatively low $\Gamma$ values, turning into a much steeper slope, above the 
transition mass-loss rate, where $\dot{M} \propto \Gamma^5$ (see also Vink 2006). This steeper slope
is not yet adopted into most stellar evolutionary models for VMS (e.g. Yusof et al. 2013; Koehler 
et al. 2015), although Chen et al. (2015) have made first attempts in this direction.

Figure 4 involves a set of theoretical computations for mass-loss rates at high Eddington factor $\Gamma$, relevant
for VMS, but they have yet to be tested against observations. In order to test the transition from 
optically thin to optically thick winds, it is important to sample the upper IMF, which is only possible for the 
most massive clusters in the Local Universe. In the context of the VLT-Flames Tarantula Survey (VFTS; Evans et al. 2011), 
we have analyzed an unprecedentedly large sample (of 62) of the most massive Of-Of/WN- WNh stars in the 30 Doradus region of the Large Magellanic Cloud (LMC) using 
CMFGEN. The results are presented in Fig.\,5 and in Bestenlehner et al. (2014). 
The most notable aspect of our empirical work is that
we do indeed {\it confirm} the presence of a kink at a critical $\Gamma$ value. Furthermore, the slope in the 
mass-loss relation of the upper $\Gamma$ range was found to be 5, also in perfect agreement with our theoretical 
Monte Carlo models (Vink 2006; Vink et al. 2011). 

\begin{figure*}
\begin{center}
\includegraphics[width=\textwidth]{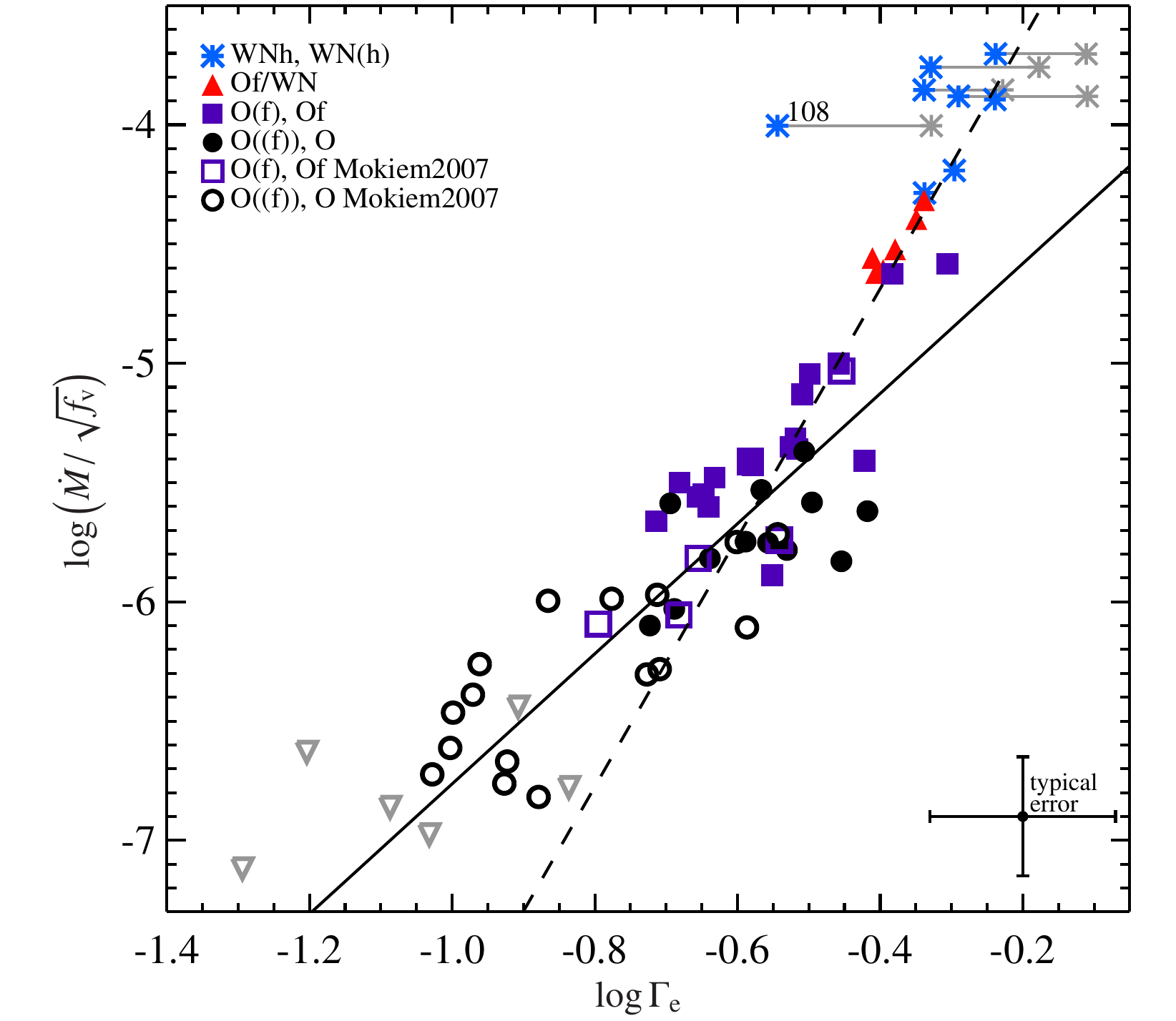}
\caption{Unclumped $\log \dot{M}$ (divided by the inverse of the 
   square-root of the clumping factor) 
   versus~$\log \Gamma_{\rm e}$ from Bestenlehner et al. (2014). 
   Solid line: $\dot{M}-\Gamma_{\rm}$ relation for O stars. The
  different symbols indicate stellar sub-classes. 
  Dashed line: the steeper slope
  of the Of/WN and WNh stars. The {\it kink} occurs at $\log
  \Gamma_{\rm e} = -0.58$. The grey asterisks indicate the
    position of the stars with $Y > 0.75$ under the assumption of
    core He-burning. The grey upside down triangles are stars from
  Mokiem et al. (2007) which have only an upper limit in $\dot{M}$ and
  are excluded from the fit.}
\label{f:mdot_gamma_obs}
\end{center}
\end{figure*}

It should be noted that this steep increase of $\dot{M}$ above a critical $\Gamma$ value cannot be explained 
by CAK theory, where the steep increase only takes place at $\Gamma$ values extremely close to 1, unless one 
were to chance the CAK-parameter $\alpha$ at the kink, but to all intents and purposes, a change in $\alpha$ at the kink 
versus a kink in its own right, represent a similar physical effect, due to the increased optical depth.

Finally, we should note that despite the qualitative success regarding (i) the presence of a kink, and (ii) the 
value of the steep slope (of 5) at the high $\Gamma$ end, the location of the kink in terms of the exact $\Gamma$ value 
(and related to different opacity treatments, and derivations of the $\Gamma$ factor for (in)homogeneous stellar evolution
models) implies that there is still quantitative work ahead of us in establishing accurate empirical and theoretical mass-loss 
rates for VMS. 

A similar uncertainty exists for massive stars in evolved evolutionary phases, e.g. 
in the LBV and classical WR phase, where the $\Gamma$ factor might be substantial too.

\section{LBVs: Envelope Inflation and Bi-stability Jumps}

\begin{figure}
\begin{center}
\includegraphics
  [width=\textwidth]{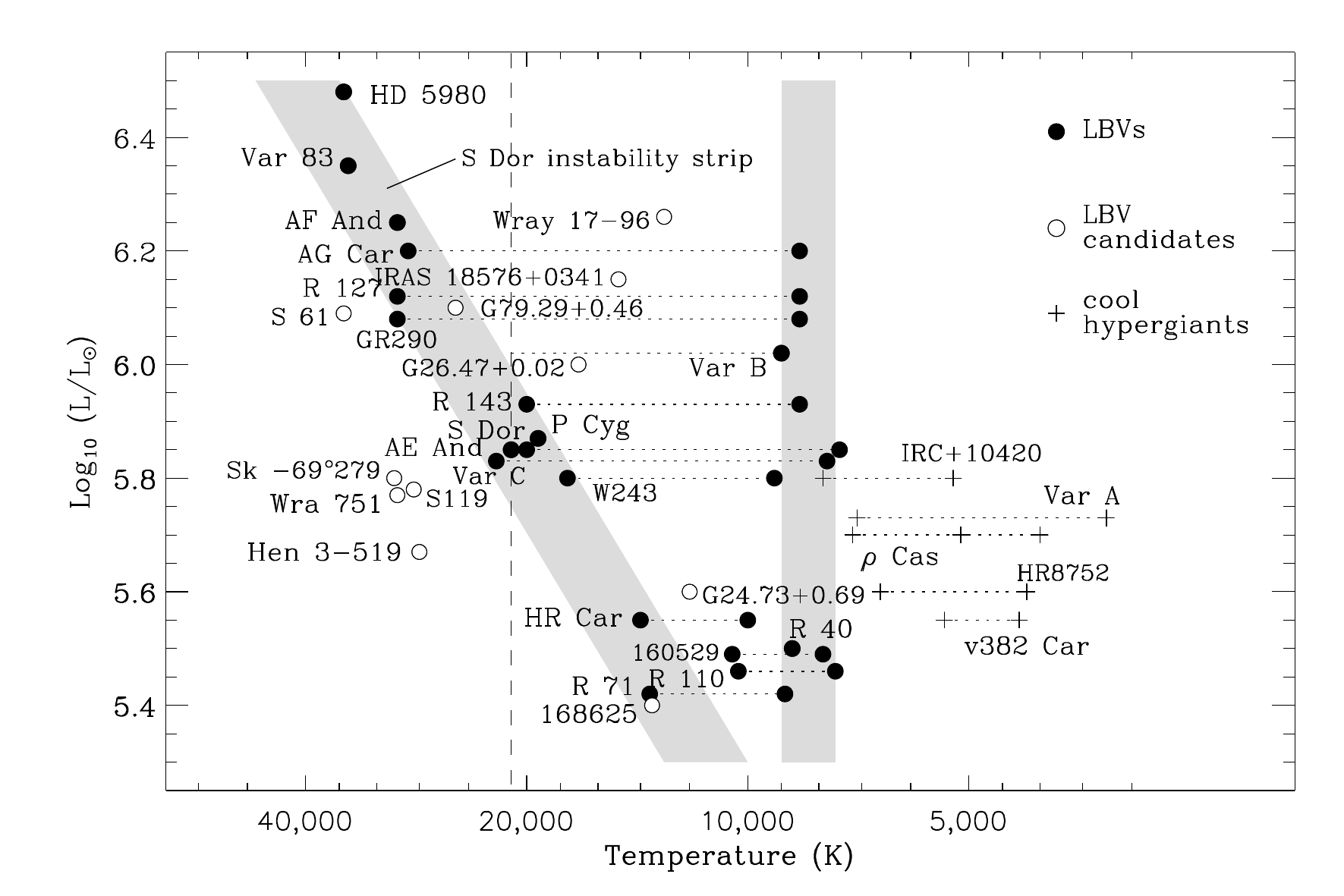}
\caption{The location of the LBVs (black circles) and candidates (open circles) in 
the Hertzsprung-Russell diagram. The cool yellow hypergiants are indicated with 
pluses. The slanted and vertical grey bands represent visual minimum and maximum 
respectively. The dashed vertical line at 21\,000K indicates the location of first 
the bi-stability jump. The figure has been adapted from Humphreys \& Davidson (1994); Smith et al. (2004); 
and Vink (2012).}
\end{center}
\end{figure}

Luminous Blue Variables (or LBVs) are objects that change spectral type (from early B to late A/early F) on timescales 
of years to decades (Humphreys \& Davidson 1994; Fig.\,6). This physical phenomenon with associated $\Delta V$ of $\sim1-2$ magnitudes is 
usually referred to as the ``S Dor'' cycle\footnote{In contrast to the giant eruption LBVs like P\,Cygni and $\eta$\,Car where $\Delta V = 5$, or more}. 
The reason for the apparent changes in effective temperature has not been
fully understood, but the leading contender is thought to be envelope inflation (Gr\"afener et al. 2012, Vink 2012; Sanyal et al. 2015). 
The typical structure of a massive star that is subject to envelope inflation is depicted in Fig.\,7.
This core-halo structure (Ishii et al. 1999; Petrovic et al. 2005; Yungelson et al. 2008; Gr\"afener et al. 2012) includes 
a very low density envelope beneath a thin shell (associated with the density inversion in Fig.\,7).

\begin{figure}
\begin{center}
\includegraphics
  [width=\textwidth,angle=90]{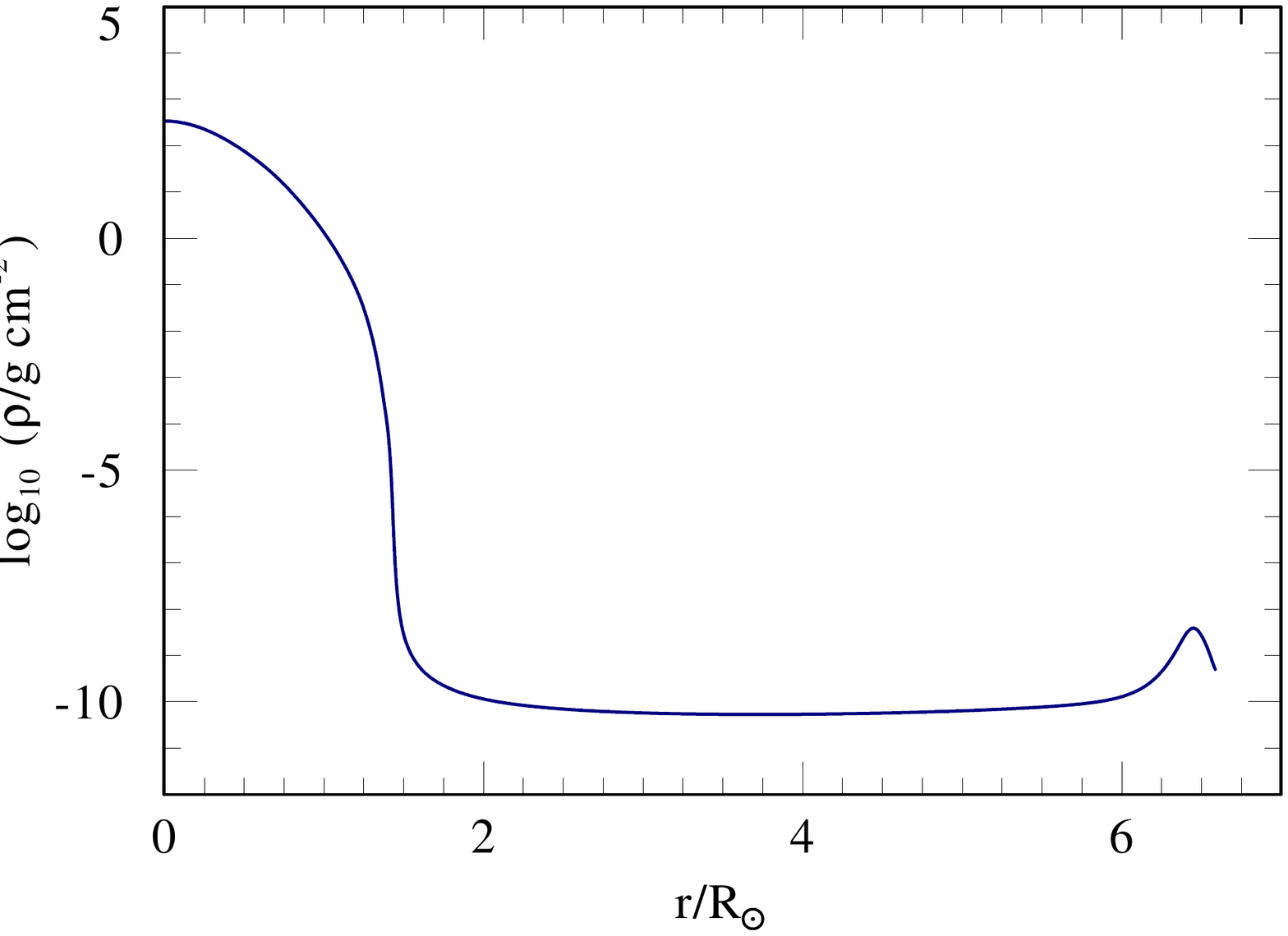}
\caption{The peculiar density structure of a massive star with an inflated envelope. The model
has a very extended low-density envelope over many stellar radii (Gr'\"afener et al. 2012).}
\end{center}
\end{figure}

\begin{figure}
\begin{center}
\includegraphics[width=\textwidth]{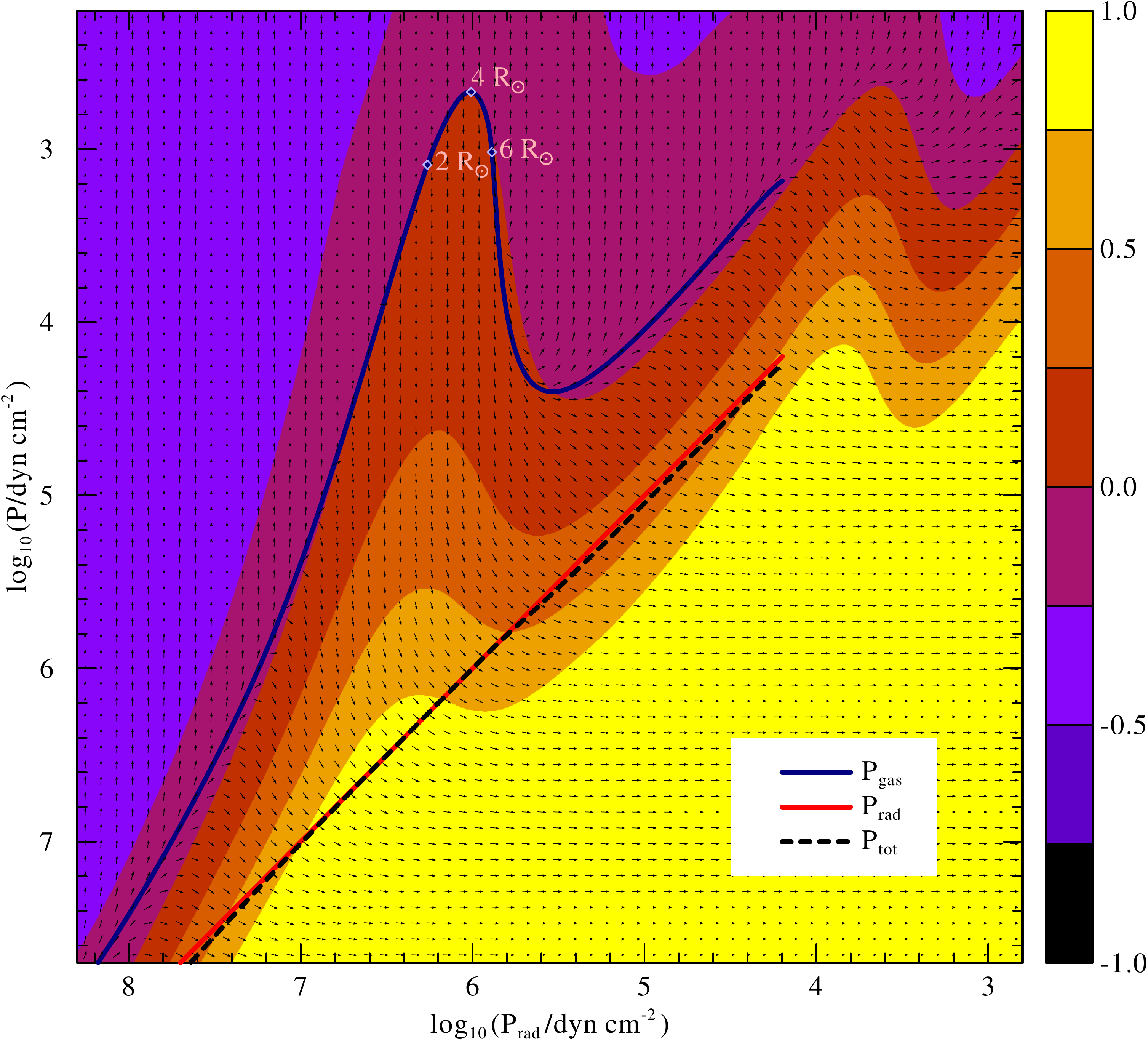}
\caption{Envelope solution in the $P_{\rm rad}$--$P_{\rm gas}$
    plane. The colours indicate the logarithm of the Eddington factor
    $\Gamma$, for given $P_{\rm rad}$ and $P_{\rm gas}$, according to
    the OPAL opacity tables -- for a pure He model
    with $23\,M_\odot$ (Gr\"afener et al. 2012). The envelope solution almost precisely follows
    a path with $\Gamma = 1$. The radii of 2, 4, and $6\,R_\odot$
    within the inflated envelope are indicated. The arrows indicate the
    slope of the solution.}
  \label{fig:gamma}
\end{center}
\end{figure}

Whilst models including the OPAL (Iglesias \& Rogers 1996) opacities showed the presence of an inflated envelope in stellar structure
and evolution calculations, it was not until Gr\"afener et al. (2012) presented an in-depth analysis of  
envelope inflation around the Fe opacity peak, that an analytic formula for radius extension was derived:

\begin{equation}
\frac{R_{\rm out}}{R_{\rm in}}~=~\frac{1}{1-W}~~~~~~~~W~=~\frac{\Delta P_{\rm rad}R_{\rm in}}{GM\rho_{\rm mean}}
\end{equation}
The formula indicates that envelope extension can be understood in terms of the width of the Fe opacity peak 
in terms of $\Delta$ $P_{\rm rad}$ divided by potential energy of the envelope (see Fig.\,8). 

If and when an LBV inflates its envelope, presumably due to its approach of the Eddington limit, it is expected to cross 
the upper Hertzsprung-Russell diagram (HRD), thereby encountering certain discontinuities 
in mass-loss rates at the first (21\,000; Pauldrach \& Puls 1990; Lamers et al. 1995; Vink et al. 1999; Petrov et al. 2016) and 
second (10\,000K; Lamers et al. 1995; Vink et al. 1999; Petrov et al. 2016) bi-stability jumps, which are due to the recombination
of Fe {\sc iv} to {\sc iii} at the first jump, and Fe {\sc iii} to Fe {\sc ii} at the second jump. 

The first bi-stability jump was indicated in the HRD of  Fig.\,6 and has been relatively well-studied. 
The second bi-stability jump was only recently studied in detail (using CMFGEN models) and the results are shown in Fig.\,9.
As LBVs have recently been suggested to be the direct progenitors of the transitional supernova types IIb and IIn(arrow line) 
(Kotak \& Vink 2006; Smith 2015), with enhanced mass loss prior to explosion, there has been renewed interest in the physical mechanism for this mass loss.
To investigate the physical ingredients that may play a role in the radiative acceleration of LBVs, Petrov et al. calculated 
blue supergiant wind models with the CMFGEN non-LTE model atmosphere code over an effective temperature range between 30000 and 8800 K, studying 
and confirming the existence of both the first and second bi-stability jumps of Vink et al. (1999). 
However, Petrov et al. (2016) found them to occur at somewhat lower $T_{\rm eff}$(20\,000 and 9\,000 K), in better accord with the observed 
locations (Lamers et al. 1995), which would imply that stars may evolve towards lower $T_{\rm eff}$ before strong mass-loss is induced by 
the bi-stability jumps.
 
When the combined effects of the second bi-stability jump and the proximity to Eddington limit are accounted for, Petrov et al. found a 
dramatic increase in the mass-loss rate by up to a factor of 30 (see Fig.\,9), which may have consequences for the evolution
of massive stars prior to expiration, as can be gleaned from Fig.\,2. 

Further investigation of both bi-stability jumps is expected to lead to a better understanding of 
discrepancies between empirical modelling and theoretical mass-loss rates reported in the literature, and to 
provide key inputs for the evolution of both normal AB supergiants and LBVs, as well as their subsequent supernova type II explosions.

\begin{figure*}
\begin{center}
\includegraphics
  [width=\textwidth]{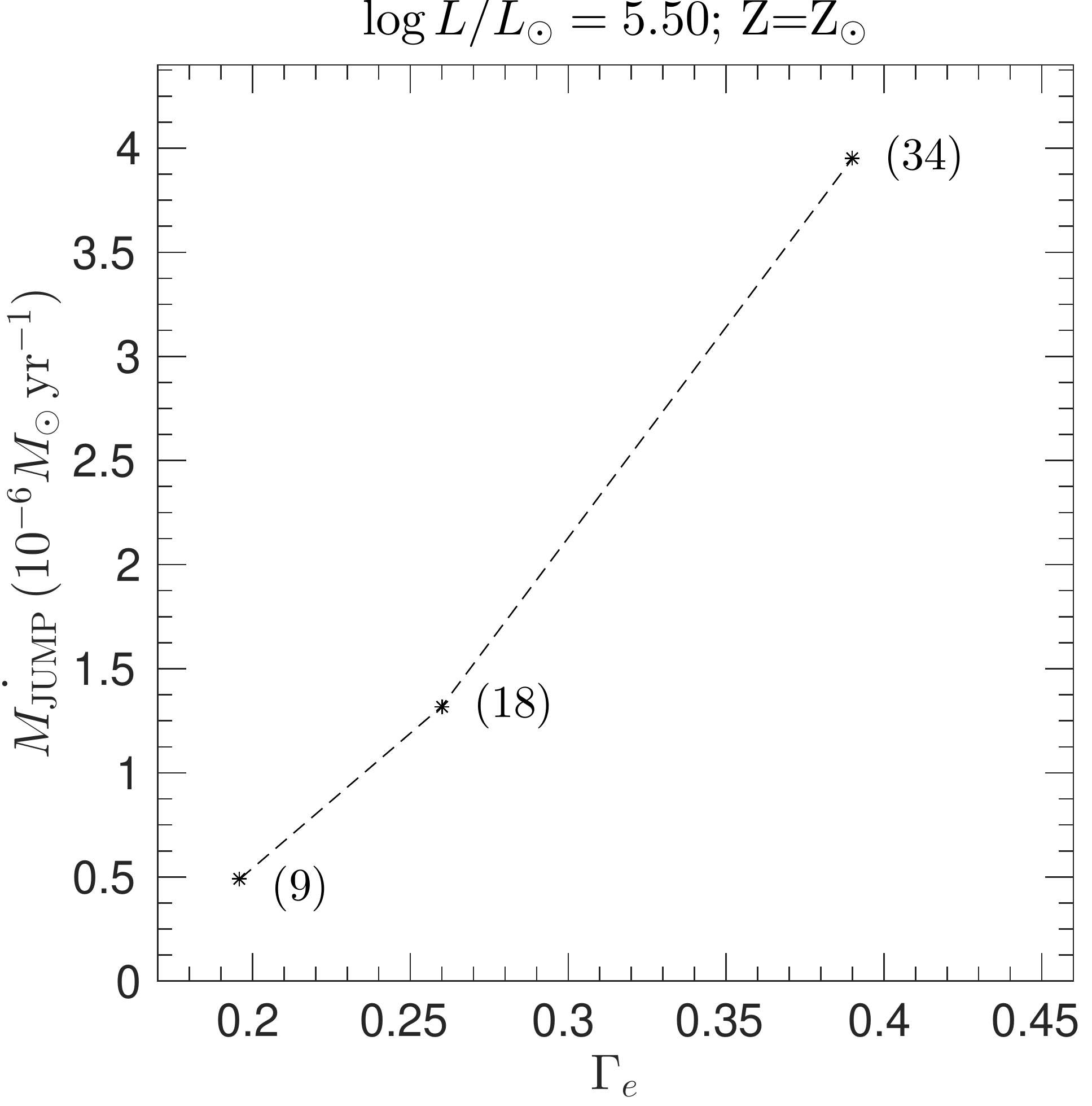}
\caption{Theoretical mass-loss models for mass-loss rates at the {\it second} bi-stability jump, as 
computed with the CMFGEN (Hillier \& Miller 1998) code by Petrov et al. (2016).}
\end{center}
\end{figure*}

\section{Conclusion}

Stellar (super)winds are not only important for correctly predicting the forwards evolution
of massive stars from the zero-age main-sequence (ZAMS) towards explosion, but also for 
the {\it backwards} identification of SN progenitors. For both aspects it is crucial to understand 
that radiation-driven wind models do not only depend on stellar luminosity (and metallicity), but 
also on effective temperature (especially at bi-stability jumps) and stellar mass 
(and/or the Eddington factor $\Gamma$). 

Only when we are able to correctly predict the mass-loss rates of evolved massive stars (at high $\Gamma$) 
as a function of all these stellar parameters will we
be able to correctly model the evolution \& fate and 
atmospheres of massive stars, including their ionizing radiation
as a function of metallicity -- and cosmic time. 

\enlargethispage{20pt}

\ack{Insert acknowledgment text here.}


\end{document}